# Excellent microwave and magnetic properties of La-Cu substituted Ba-hexaferrites prepared by ceramic process

Chuanjian Wu[1], Zhong Yu[1], Ke Sun[1,*], Alexander S. Sokolov[2,*], Rongdi Guo[1], Yan Yang[3], Chengju Yu[2], Xiaona Jiang[1], Zhongwen Lan[1], and Vincent G. Harris[2]

[1]*State Key Laboratory of Electronic Thin Films and Integrated Devices, University of Electronic Science and Technology of China, Chengdu 610054, China*

[2]*Department of Electrical and Computer Engineering, Northeastern University, Boston, MA 02115, USA*

[3]*Department of Communication and Engineering, Chengdu Technological University, Chengdu 611730, China*

We demonstrate exceptional microwave magnetic performance of crystallographycally textured barium hexaferrites prepared by conventional ceramic method. Vibrating sample magnetometer (VSM) and rectangular cavity perturbation measurements indicate that bulk $Ba_{0.8}La_{0.2}Fe_{11.8}Cu_{0.2}O_{19}$ samples exhibit a strong magnetic anisotropy field $H_a$ of 14.6kOe, the highest remanent magnetization $4\pi M_r$ of 3.96kGs, and the lowest ferromagnetic resonance (FMR) linewidth $\Delta H$ of 40 1Oe with zero applied bias field at Q-band. Broadband millimeter-wave transmittance measurements determined the complex permeability $\mu^*$ and permittivity $\varepsilon^*$. The remanent magnetization $4\pi M_r$, anisotropy field $H_a$, and effective linewidth $\Delta H_{eff}$ were also estimated based on Schlömann's theory of the complex permeability $\mu^*$. These calculated values are in good agreement with the VSM and FMR measurements.

**Introduction**

Microwave ferrites have found numerous applications in both defense and commercial systems, such as circulators or isolators[1-2]. Despite the progress made over the last few decades, these devices remain quite large and heavy because of the external permanent magnets required to provide indispensable biasing fields. On the other hand, crystallographycally textured ferrites allow for the exploitation of their built-in magnetic fields (i.e., self-bias effects), and thus, integration and miniaturization of microwave and millimeter wave (MMW) devices. Specifically, M-type barium hexaferrites ($BaFe_{12}O_{19}$ or BaM) have attracted attention for applications in monolithic microwave integrated circuits (MMIC) in virtue of their strong and tunable magnetocrystalline anisotropy field $H_a$ that acts as the effective internal field[3-4].

Comparatively low insertion losses over a broad bandwidth are a prerequisite for favorable high-frequency characteristics of self-biased MMW devices based on crystallographycally textured hexaferrites[2]. Improved remanent magnetization $4\pi M_r$ and narrowed ferromagnetic resonance (FMR) linewidths $\Delta H$ allow for the reduction of the resonance losses[5], and therefore, should be optimized along with the anisotropy field $H_a$. **Fig.1(a)** enumerates a series of published values of remanent magnetization $4\pi M_r$ and FMR linewidth $\Delta H$ of bulk barium heaferrites[6-14]. It is clear that conventionally processed bulk BaM usually exhibits a relatively low remanent magnetization $4\pi M_r$ of 1.5~3.5kGs and broad FMR linewidth $\Delta H$ of 0.5 to 2kOe. Herein we demonstrate crystallographycally textured hexaferrites with simultaneously high remanent magnetization $4\pi M_r$ and low ferromagnetic resonance (FMR) linewidth $\Delta H$.

*Corresponding author. Email addresses: ksun@uestc.edu.cn (Dr. K. Sun), sokolov.al@husky.neu.edu (Dr. A. S. Sokolov);

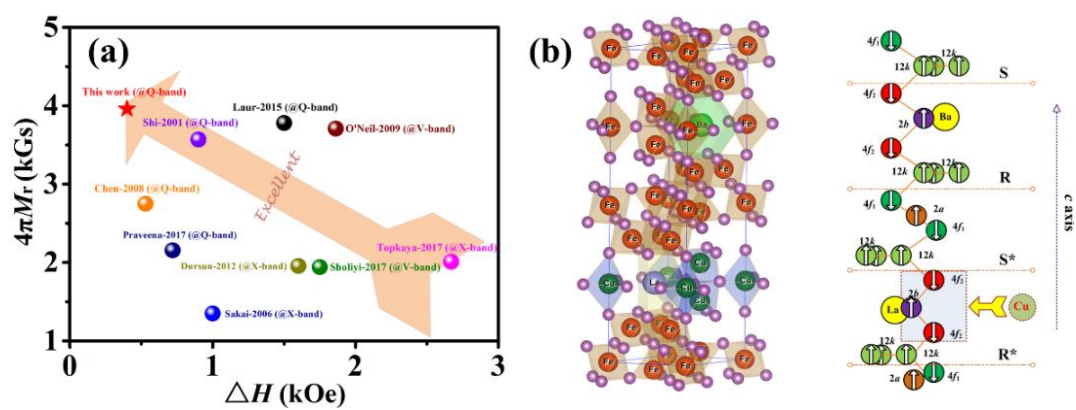

**Fig.1** Published values for remanent magnetization $4\pi M_r$ and FMR linewidth $\Delta H$ in bulk BaM and the corresponding crystalline structure of La-Cu substituted barium hexaferrites in this work

Furthermore, dynamic magnetic permeability $\mu^*$ and dielectric permittivity $\varepsilon^*$ are the two fundamental parameters that determine the response of the ferrites to high frequency magnetic and electromagnetic fields. We employed high-$Q$ resonant measurements to determine dielectric and magnetic properties at microwave frequencies[11]. The accuracy of this method deteriorates in the millimeter-wave region of the electromagnetic spectrum, therefore, the magneto-optical approach has been employed. This technique allows for the separation of the dielectric and magnetic effects in ferrites, and enables the holonomic and precise characterization of ferrites in the entire millimeter-wave range[15].

Basic operational principles of MMV devices require the thickness of crystallographycally textured ferrites to exceed (300μm)[2], therefore bulk hexaferrites remain somewhat preferable to their film counterparts. In the present work, we document the processing-structure-property relationships of La-Cu substituted Ba-hexaferrites prepared by conventional ceramic process. The crystalline orientation, microstructure, magnetic and microwave properties are investigated. The complex permeability $\mu^*$, complex permittivity $\varepsilon^*$, and effective linewidth $\Delta H_{eff}$ are determined by the magneto-optical measurements.

**Experimental Procedures**

Previous calculations[16] of the density functional theory (DFT) revealed that $Cu^{2+}$ ions are positioned at the $2b$ and $4f_2$ sites (see **Fig.1(b)**). The inclusion of $Cu^{2+}$ ions allows to adjust the microwave and magnetic properties of crystallographycally textured hexaferrites. Here in this work, $Ba_{0.8}La_{0.2}Fe_{11.8}Cu_{0.2}O_{19}$ was synthesized according to the conventional ceramic process. Analytical-grade raw materials, $BaCO_3$, $Fe_2O_3$, $La_2O_3$, and $CuO$ powders were homogeneously mixed by ball-milling in an agate jar for 12 hours. These composite mixtures were subsequently calcined at 1100~1200ºC for 2 hours, doped with 2.5wt% of $Bi_2O_3$ and 2.0wt% of $CuO$, and ball-milled again for 18 hours to achieve a mean particle size of 0.6~1.0μm. Following the ball-milling, the powder was dried, then deionized water was added to form an aqueous slurry with a 60~70wt% solids loading. Subsequently the slurries were uniaxially pressed in a strong longitudinal magnetic field (~15kOe) to produce crystallographycally textured BaM cylindrical green bodies, the $c$-axes of barium hexaferrite particles were parallel to the cylinder axes. The specimens were then sintered at 950~1050 ºC for 2 hours.

Crystallographic structure was determined by means of X-ray diffraction (XRD, Bruker D8), using a Cu$K\alpha$ radiation source. The cross-section morphology was observed using a field emission scanning electron microscope (FESEM, JEOL JSM-7800F). The bulk density $d$ was measured by Micromeritics AccuPyc 1330 pycnometer. Magnetic properties at room temperature were determined by a vibrating sample magnetometer (VSM, Quantum Design SQUID), the maximum applied magnetic fields equaled 20kOe. Ferromagnetic resonance (FMR) linewidth spectra were characterized by the $TE_{10}$ rectangular transmission cavity perturbation at Q-band (30~50GHz)[17,18]. Broadband millimeter-wave measurements were performed using a free space quasi-optical spectrometer, equipped with a series of backwards wave oscillators (BWO's) as high-power, tunable sources of coherent radiation within the frequency range of 30~90GHz.

**Results and discussion**

**Fig.2(a)** shows the respective X-ray diffraction patterns of crystallographycally textured barium hexaferrites. All diffraction peaks have been indexed according to the standard powder diffraction pattern of the JCPDF card (No.43-0002) for $BaFe_{12}O_{19}$, corresponding to the space group $P6_3/mmc$. The peaks perpendicular to the sample plane of (006), (008), (0014) have the highest intensity, however other weak peaks are also present, namely (107), (206), and (220). As a reliable way to evaluate the degree of crystalline orientation, the Lotgering factor $f_L$ was calculated[19]

$$f_L = \frac{\sum I(00l)/\sum I(hkl) - \sum I_0(00l)/\sum I_0(hkl)}{1 - \sum I_0(00l)/\sum I_0(hkl)} \quad (1)$$

where $\sum I(00l)/\sum I(hkl)$ is calculated based on the experimental data for crystallographycally textured samples, and $\sum I_0(00l)/\sum I_0(hkl)$ is based on the XRD data on the standard card. The evaluated degree of crystalline orientation $f_L$ for $Ba_{0.8}La_{0.2}Fe_{11.8}Cu_{0.2}O_{19}$ samples reaches 83%, indicating that textured barium hexaferrites exhibit a strong $c$-axis preferred orientation.

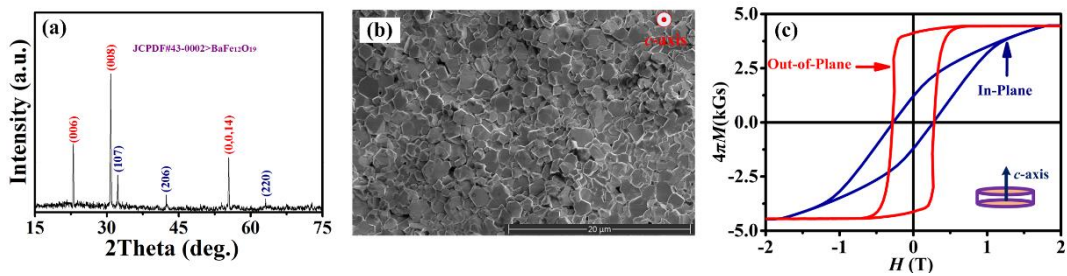

**Fig.2** The crystallographic, microstructural and magnetic properties of textured hexaferrites.

The cross-section morphology of crystallographycally textured barium hexaferrites is presented in **Fig.2(b)**. It can be observed that the textured $Ba_{0.8}La_{0.2}Fe_{11.8}Cu_{0.2}O_{19}$ sample demonstrates a microstructure consisting of hexagonal particles compactly aligned parallel to $c$-axis. It is crucial to control the grain size during the preparation of textured samples with high remanent magnetization $4\pi M_r$. The grain size ranges from 1 to 2μm, and the average grain size is 1.2μm, which lies within the estimated single domain critical size range (1.0-1.5$\mu m$)[20]. Additionally, the density $d$ of as-sintered $Ba_{0.8}La_{0.2}Fe_{11.8}Cu_{0.2}O_{19}$ specimen was measured to be 97.5% of the theoretical density (~5.28g/cm$^3$)[21]. As indicated previously in Ref.[22 and 23], bismuth copper oxides, generated in the reaction of $Bi_2O_3$ and CuO, could establish a liquid-channel structure located at the grain boundaries. It is speculated that this structure promotes the densification and grain alignment during sintering. Yet, some pores are still visible in **Fig.2(b)** after sintering, and inevitably contribute to the broadening of the FMR linewidth.

**Fig.2(c)** illustrates static magnetic hysteresis loops ($M$ versus $H$) of the textured $Ba_{0.8}La_{0.2}Fe_{11.8}Cu_{0.2}O_{19}$ sample. The kink in the out-of-plane loop evolves slowly, thus indicating the strong $c$-axis orientation in the textured $Ba_{0.8}La_{0.2}Fe_{11.8}Cu_{0.2}O_{19}$ sample. To describe obvious divergences between the in-plane and out-of-plane loops, the anisotropy field $H_a$ was obtained by the SPD (singular point detection) method[14], and the anisotropy constant $K_1$ calculated in terms of the basic equation $H_a=2K_1/M_s$. All representative magnetic properties are summarized in **Table 1**. It was found that textured $Ba_{0.8}La_{0.2}Fe_{11.8}Cu_{0.2}O_{19}$ samples show a strong uniaxial anisotropy with anisotropy field $H_a$ of 14.6kOe and anisotropy constant $K_1$ of $2.52\times10^5$J/m$^3$. Note that observed remanent magnetization $4\pi M_r$ of 3.96kOe is superior to the previously reported results for bulk BaM (see **Fig.1(a)**). This improvement in $4\pi M_r$ is mainly due to the uniform crystalline grains and higher sintering density. Such a strong anisotropy $H_a$ and high remanent magnetization $4\pi M_r$ enable the MMW devices to be fully operational without biasing magnets[6,7,12].

**Table 1** Physical, magnetic, and microwave properties of textured barium hexaferrites

| $\rho$(g/cm$^3$) | $4\pi M_s$ (kGs) | $4\pi M_r$ (kGs) | $H_c$ (Oe) | $H_a$(kOe) | $K_1(\times10^5$J/m$^3$) | $\Delta H$ (Oe) |
|---|---|---|---|---|---|---|
| 5.15 | 4.45 | 3.96($\perp$) | 2813($\perp$) | 14.6 | 2.52 | 379@1kOe |
|  |  | 1.22($//$) | 2698($//$) |  |  | 401@0kOe |

The frequency response of the resonance absorption at Q-band was measured with the external biasing fields $H$ of 0 and 1kOe, applied parallel to the cylinder axes. As depicted in **Fig.3**, the fitting of the resonance absorption spectrum was characterized by a Lorentzian function consisting of symmetric and asymmetric components[24]:

$$P = L_{sym}\frac{\Delta f^2}{(f-f_{res})^2+\Delta f^2} + D_{asym}\frac{\Delta f(f-f_{res})}{(f-f_{res})^2+\Delta f^2} \qquad (2)$$

where $\Delta f$ denotes the half width at half maximum (HWHM), and $f_{res}$ is ferromagnetic resonance frequency. $L_{sym}$ and $D_{asym}$ are respectively symmetric and asymmetric contributions to the frequency. The corresponding FMR linewidth $\Delta H$ of the textured $Ba_{0.8}La_{0.2}Fe_{11.8}Cu_{0.2}O_{19}$ sample was evaluated according to the equation $\Delta H=2\times\Delta f/\gamma$ and listed in **Table 1**. The strong absorption zones reside in the vicinity of the respective center frequencies of 44.0GHz and 46.3GHz for crystallographycally textured hexaferrites at $H$=0 and 1kOe. According to Kittel formula[14], the anisotropy field $H_a$ obtained from FMR measurements was extrapolated to be 15.1±0.2kOe, which is very close to the one derived from the SPD method. It is noteworthy that the crystallographycally textured $Ba_{0.8}La_{0.2}Fe_{11.8}Cu_{0.2}O_{19}$ sample exhibited a narrow FMR linewidth $\Delta H$ of 40Oe at zero bias field, an improvement of over 20% as compared to the previous values for bulk BaM (see **Fig.1(a)**).

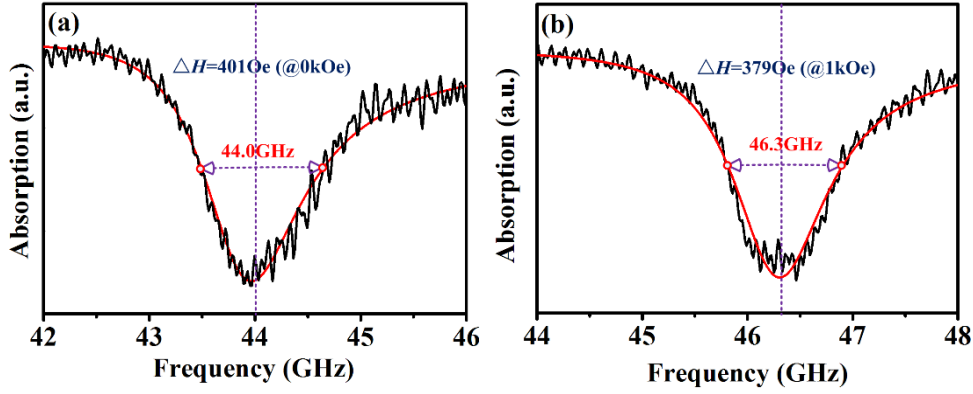

**Fig.3** The dependence of the frequency response of the resonance absorption in a crystallographycally textured $Ba_{0.8}La_{0.2}Fe_{11.8}Cu_{0.2}O_{19}$ sample: (a) at 0Oe; (b) at 1kOe.

In order to probe into the reasons of the low FMR linewidth $\Delta H$, it is essential to estimate the contributions of different factors to the FMR linewidth $\Delta H$. In polycrystalline ferrites, the total FMR linewidth $\Delta H$ depends crucially on the superposition of intrinsic and extrinsic contributions.[25]

$$\Delta H = \Delta H_i + \Delta H_a + \Delta H_p \qquad (3)$$

where $\Delta H_i$ is the intrinsic linewidth; Karim et.al.[26] speculated that barium hexaferrites poses an intrinsic linewidth of 0.3~0.4Oe/GHz. $\Delta H_a$ and $\Delta H_p$ relate to the crystalline anisotropy and porosity induced linewidth broadening contributions. Approximate estimations of $\Delta H_a$ and $\Delta H_p$ were proposed by Schlömann,[27,28] based on the independent grain approaches with large anisotropy ($H_a \gg 4\pi M_s$):

$$\begin{cases} \Delta H_a \approx 0.87 H_a \\ \Delta H_p \approx 1.5(4\pi M_s)P \end{cases} \qquad (4)$$

where $P$ corresponds to the porosity. This evaluation indicates that 80%~85% of the total FMR linewidth is determined by the porosity contribution. In other words, a relatively high density of 97.5% could allow for an even lower FMR linewidth $\Delta H$ in textured $Ba_{0.8}La_{0.2}Fe_{11.8}Cu_{0.2}O_{19}$ samples.

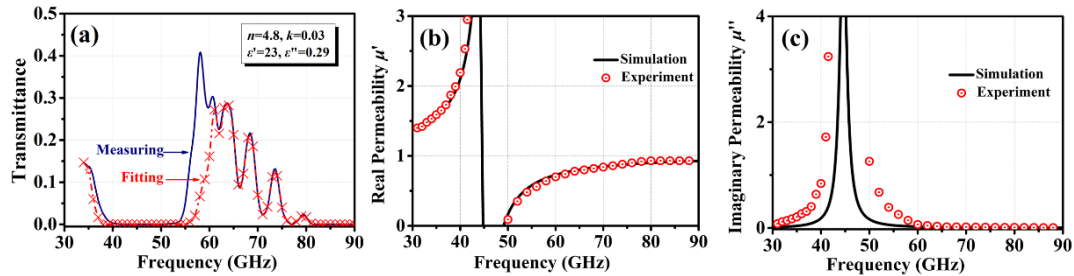

**Fig.4** Millimeter wave transmittance spectra, and real and imaginary parts of magnetic permeability in crystallographycally textured $Ba_{0.8}La_{0.2}Fe_{11.8}Cu_{0.2}O_{19}$ sample as a function of frequency.

The millimeter-wave transmission spectrum of textured $Ba_{0.8}La_{0.2}Fe_{11.8}Cu_{0.2}O_{19}$ samples was recorded in **Fig.4**. A relatively wide absorption zone centered at about 44.3GHz is thought to be a natural ferromagnetic resonance[29]. Nevertheless, the actual width of resonance could not be obtained from the absorption line, which is ascribed mainly to the broadening experimental width induced by the saturation of the absorption line. At frequencies below and above this zone, a progressive decline of the transmissivity and some oscillations are observed. Such variations in the microwave or millimeter ranges are completely associated with magnetic permeability[30]. By integrating the relationship between the transmittance and reflectance spectra, and also refractive and absorption indexes, we could obtain[31]

$$\begin{cases} T = E \dfrac{(1-R)^2}{(1-RE)^2 + 4RE \sin^2(2\pi n v d/c)} \\ R = \dfrac{(n-1)^2 + k^2}{(n+1)^2 + k^2} \\ E = \exp(-\dfrac{4\pi k d f}{c}) \\ n + ik = \sqrt{\varepsilon^* \mu^*} \end{cases} \quad (5)$$

where $T$, $R$, $n$, $k$, $\mu^*$, and $\varepsilon^*$ are the transmittance, reflectance, refractive index, absorption index, complex permeability, and complex dielectric permittivity, respectively. The dielectric permittivity $\varepsilon^*$ typically exhibits a very weak frequency dependence in the millimeter range, and therefore was reasonably assumed to be constant. By fitting the transmission spectrum far from the absorption zone, it was deduced that textured $Ba_{0.8}La_{0.2}Fe_{11.8}Cu_{0.2}O_{19}$ samples had a real permittivity $\varepsilon'$ of 23, and imaginary permittivity $\varepsilon''$ of 0.29. The complex magnetic permeability was evaluated based on the magnetocrystalline anisotropy $H_a$ and remanent magnetization $4\pi M_r$[32]:

$$\mu^*/\mu_0 = sqrt\{[(f_a^* + f_{4\pi M_r})^2 - f^2]/[(f_a^*)^2 - f^2]\} \quad (6)$$

where $f_a^* = (\gamma/2\pi)H_a + jfG$, and $f_{4\pi M_r} = 2\gamma(4\pi M_r)$ with $\gamma$ being the gyromagnetic ratio, and $G$ being the damping parameter. The fitting of the experimentally observed spectrum was performed by using Eq.(6). It is noteworthy that a significant discrepancy between the experimental and simulated results for the resonance absorption zone is attributed to the considerable contribution of "non-intrinsic" relaxation. The best-fit dissipation leads to $H_a$=15.4kOe, $4\pi M_r$=4.02kGs, and $G$=0.0011 far from the resonance. These values of $H_a$ and $4\pi M_r$ are in good agreement with the ones obtained from the VSM and FMR measurements. According to the equation $\Delta H_{eff} = 2\pi G \cdot f/\gamma$[29], the effective linewidth $\Delta H_{eff}$ equals 108Oe.

This narrow $\Delta H_{eff}$ is beneficial for the design of MMW devices that operate far from the resonance frequency.

**Conclusions**

In summary, crystallographycally textured $Ba_{0.8}La_{0.2}Fe_{11.8}Cu_{0.2}O_{19}$ ferrites were successfully fabricated according to the conventional oxide ceramic process. These ferrites exhibit a strong crystallographic *c*-axis alignment, high sintering density *d* of 5.15g/cm$^3$, strong anisotropy field $H_a$ of 14.6kOe, and the highest remanent magnetization $4\pi M_r$ of 3.96kGs. The corresponding FMR linewidth $\Delta H$ with biasing fields *H* of 0 and 1kOe at Q- band were measured to be 401Oe and 379Oe, respectively. Moreover, the magneto-optical measurements allowed to determine the complex dielectric permittivity $\varepsilon^*$, permeability $\mu^*$, and effective linewidth $\Delta H_{eff}$. While the permeability $\mu^*$ strongly depends on the frequency, the real ($\varepsilon'$) and imaginary parts ($\varepsilon''$) of the dielectric permittivity $\varepsilon^*$ and effective linewidth $\Delta H_{eff}$ respectively equaled 23, 0.29, and 108Oe far from the resonance. These results indicate that crystallographycally textured $Ba_{0.8}La_{0.2}Fe_{11.8}Cu_{0.2}O_{19}$ ferrites can be employed in self-biased and low-loss MMW devices, such as circulators or isolators.

**Acknowledgements**

We acknowledge Lt. K. A. Korolev, Prof. K. N. Kocharyan and M. N. Afsar from Tufts University for their helps with the magneto-optic measurements and discussions. This work is supported by the China Scholarship Council and National Natural Science Foundation of China under Grant No. 51772046.

**References**

[1] Z. Su, H. Chang, X. Wang, A. S. Sokolov, B. Hu, Y. Chen, and V. G. Harris, Appl. Phys. Lett. **105**, 062402 (2014).

[2] V. G. Harris, Z. Chen, Y. Chen, S. Yoon, T. Sakai, A. Gieler, A. Yang, and Y. He, J. Appl. Phys. **99**, 08M911 (2006).

[3] D. Chen, G. Wang, Z. Chen, Y. Chen, Y. Li, and Y Liu, Mater. Lett. **189**, 229 (2017).

[4] Y. Song, J. Das, Z. Wang, W. Tong, and C. E. Patton, Appl. Phys. Lett. **93**, 172503 (2008).

[5] X. Zuo, H. How, S. Somu, and C. Vittoria, IEEE T. Magn. **39**, 3160 (2003).

[6] V. Laur, G. Vérissimo, P. Quéffélec, A. Farhat, H. Alaaeddine, J. Reihs, E. Laroche, G. Martin, R. Lebourgeois, and J. Ganne, IEEE MTT-S International **7166760**, 1 (2015).

[7] B. K. O'Neil, and J. L. Young, IEEE T. Microw. Theory **57**, 1669 (2009).

[8]S. Dursun, R. Topkaya, N. Akdogan, and S. Alkoy, Ceram. Int. **38**, 3801 (2012).

[9]R. Topkaya, Appl. Phys. A-Mater. **123**, 488 (2017).

[10]O. Sholiyi, and J. Williams, Mater. Res. Express. **046107**, 1 (2014).

[11]K. Praveena, K. Sadhana, H. Liu, and M. Bououdina, J. Magn. Magn. Mater. **426**, 604 (2017).

[12]P. Shi, H. How, X. Zuo, S. D. Yoon, S. A. Oliver, and C. Vittoria, IEEE T. Magn. **37**, 2389 (2001).

[13]T. Sakai, Y. Chen, C. N. Chinnasamy, C. Vittoria, and V. G. Harris, IEEE T. Magn. **42**, 3353 (2006).

[14]Y. Chen, M. J. Nedoroscik, A. L. Ceiler, C. Vittoria, and V. G. Harris, J. Am. Ceram. Soc. **91**, 2952 (2008).

[15]K. N. Kocharyan, M. N. Afsar, and I. I. Tkachov, IEEE T. Microw. Theory. **47**, 2636 (1999).

[16]C. Wu, Z. Yu, Y. Yang, K. Sun, J. Nie, Y. Liu, X. Jiang, and Z. Lan, J. Alloy. Compd. **664**, 406 (2016).

[17]A. S. Sokolov, M. Geiler, and V. G. Harris, Appl. Phys. Lett. **108**, 172408 (2016).

[18]A. S. Sokolov, P. Andalib, Y. Chen, and V. G. Harris, IEEE T. Microw. Theory. **64**, 3772 (2016).

[19]P. Chang, L. He, D. Wei, and H. Wang, J. Eur. Ceram. Soc. **36**, 2519 (2016).

[20]R. C. Pullar, Prog. Mater. Sci. **57**, 1191 (2012).

[21]G. Tan, and X. Chen, J. Magn. Magn. Mater. **327**, 87 (2013).

[22]C. Wu, Z. Yu, G. Wu, K. Sun, Y. Yang, X. Jiang, R. Guo, and Z. Lan, IEEE T. Magn. **51**, 1 (2015).

[23]C. Mallika, and O. M. Sreedharan, J. Alloy. Compd. **216**, 47 (1994).

[24]Z. Zhang, L. Bai, X. Chen, H. Guo, X. L. Fan, D. S. Xue, D. Houssameddine, and C. M. Hu, Phys. Rev. B **94**, 064414 (2016).

[25]R. Guo, Z. Yu, Y. Yang, X. Jiang, K. Sun, C. Wu, Z. Xu, and Z. Lan, J. Alloy. Comp. **589**, 1 (2014).

[26]R. Karim, S. D. Ball, J. R. Truedson, and C. E. Patton, J. Appl. Phys. **73**, 4512 (1993).

[27]E. F. Schlömann, Phys. Rev. **182**, 632 (1969).

[28]E. F. Schlöemann, J. Appl. Phys. **38**, 5027 (1967).

[29]K. D. McKinstry, C. E. Patton, M. A. Wittenauer, M. Sankararaman, J. Nyenhuis, F. J. Friedlaender, H. Sato, and A. Schindler, IEEE Trans. Magn. **25**, 3482 (1989).

[30]J. F. Dillon, E. M. Gyorgy, and J. P. Remeika, J. Appl. Phys. **41**, 1211 (1970).

[31]G. V. Kozlov, S. P. Lebedev, A. A. Mukhin, A. S. Prokhorov, I. V. Fedorow, A. M. Balbashow, and I. Y. Parsegov, IEEE Trans. Magn. **29**, 3443 (1993).

[32]E. F. Schlöemann, J. Appl. Phys. **41**, 204 (1970).